\documentclass[authoryear,round,12pt]{article}
\usepackage{color}
\usepackage{pslatex}
\usepackage{graphicx}
\usepackage{setspace}
\usepackage{natbib}
\usepackage{amsmath}
\usepackage{subfigure}
\newcommand{\beq}{\begin{equation}}
\newcommand{\eeq}{\end{equation}}
\newcommand{\myabstract}{We present results of a forecast initiated
  following assimilation of observations for week Week 50 (i.e. the
  forecast begins December 16, 2012) of the 2012-2013 influenza season
  for municipalities in the United States.  The forecast was made on
  December 21, 2012.  Results from forecasts initiated the three
  previous weeks (Weeks 47-49) are also presented.  Also results from
  forecasts generated with the SIRS model without absolute humidity
  forcing (no AH) are shown.}

\newcommand{\myacknow}{Funding was provided by US NIH grant GM100467
  (JS, AK, ML), as well as NIEHS Center grant ES009089 (JS) and
  the RAPIDD program of the Science and Technology Directorate, US
  Department of Homeland Security (JS).  The content is solely the
  responsibility of the authors and does not necessarily represent the
  official views of the National Institute Of General Medical
  Sciences, National Institutes of Health, or Department of Homeland
  Security.}

\begin{document}
%
%
\title{\textbf{\large{Week 50 Influenza Forecast for the 2012-2013
      U.S. Season}}}
%
%
\author{\textsc{Jeffrey Shaman}
                                \thanks{\textit{Corresponding author address:} 
                                Jeffrey Shaman, Department of
                                Environmental Health Sciences, Mailman
                                School of Public Health, Columbia
                                University, 722 West 168th Street,
                                Rosenfield Building, Room 1104C, New
                                York, NY 10032. 
                                \newline{E-mail:
                                  jls106@columbia.edu}}\quad\textsc{}\\
\centerline{\textit{\footnotesize{Department of Environmental Health Sciences,
    Mailman School of Public Health, Columbia University, New York, New York}}}
\and
\centerline{\textsc{Alicia Karspeck}} \\
\centerline{\textit{\footnotesize{Climate and Global Dynamics
      Division, National Center for Atmospheric Research, Boulder, Colorado}}}
\and 
\centerline{\textsc{Marc Lipsitch}} \\
\centerline{\textit{\footnotesize{Center for Communicable Disease
      Dynamics, Harvard School of Public Health, Harvard University,
      Boston, Massachussetts}}}
}

\maketitle

{
\begin{abstract}
\myabstract
\end{abstract}
}

\section{Background}
\label{sec:retrofore}

Forecast skill is calculated for individual cities, as well as for
census divisions (regions) and all cities in aggregate, from
retrospective forecasts made for the 2003-2004 through 2011-2012
seasons, excluding the pandemic years 2008-2009 and 2009-2010 ( which
will need to be handled separately in the future).  The forecast
methods are similar to those described in
\cite{Shaman-Karspeck-2012:forecasting}.  Based on the relationship
between prediction accuracy and ensemble spread of these retrospective
forecasts we can assign calibrated confidences to our current
predictions.

\begin{figure}[tbh]
\noindent\includegraphics[width=18pc,angle=0]{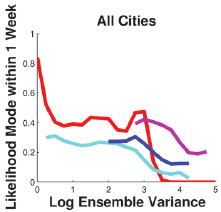}
\noindent\includegraphics[width=18pc,angle=0]{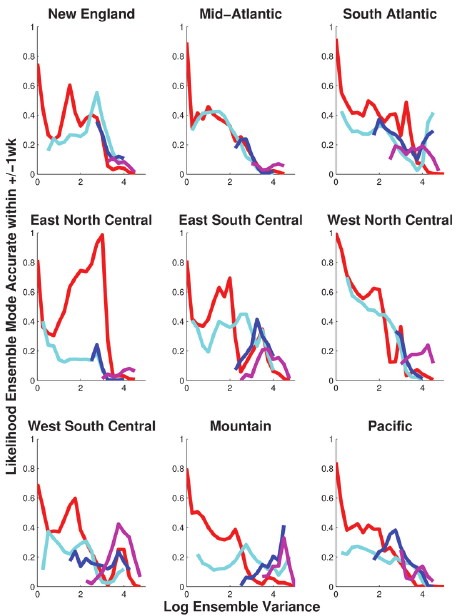}\\
\caption{Plot of ensemble mode forecast accuracy versus ensemble
  spread measured as log ensemble variance +1.  Left) 114 cities in
  aggregate.  The runs are binned in
  increments of 0.25 units and stratified by forecast lead time: 1-3
  weeks (red), 4-6 weeks (cyan), 7-9 weeks (blue), 10+ weeks
  (magenta).  Right) Same as left, but the 114 cities grouped by census region.}
\label{fig:allcities_mode_allclim}
\end{figure}

Figure \ref{fig:allcities_mode_allclim} shows the results for all
cities in aggregate using climatological AH and a factor of 5 mapping.
Overall the relationship is informative; however, for all lead times
there is a basic plateau of skill once the ensemble log variance drops
below 2.5 to 3 weeks$^2$.

When the cities are grouped by region, there is some heterogeneity.
Some regions (e.g. West North Central) show marked improvement of
forecast accuracy/skill with decreasing spread across all lead times.
Other regions show much more limited skill--the Mountain region only
has skill at 1-3 weeks, and the East North Central has problems at 1-3
weeks.  
 
\section{2012-2013 Forecast}
\label{sec:actualfore}

This record is part of an evolving series of real-time forecasts
developed during the 2012-2013 influenza season for the United
States.  Additional documentation of earlier forecasts for this season
have also been posted \citep{Shaman-Karspeck-Lipsitch-2012:week49}.

\subsection{Week 50 Forecast}
\label{subsec:actual50}

\begin{figure}[tbh]
\noindent\includegraphics[width=25pc,angle=0]{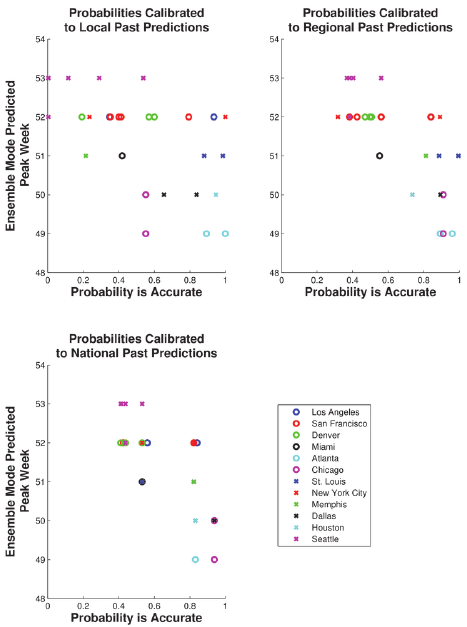}\\
\caption{Ensemble mode peak week predictions initiated December 16,
  2012, following assimilation of Week 50 observations, for 12 cities
  plotted as a function of probability/confidence calibrated from
  historical city, regional and national prediction accuracy.}
\label{fig:select_wk50fore_cal} 
\end{figure}

\begin{figure}[tbh]
\noindent\includegraphics[width=18pc,angle=0]{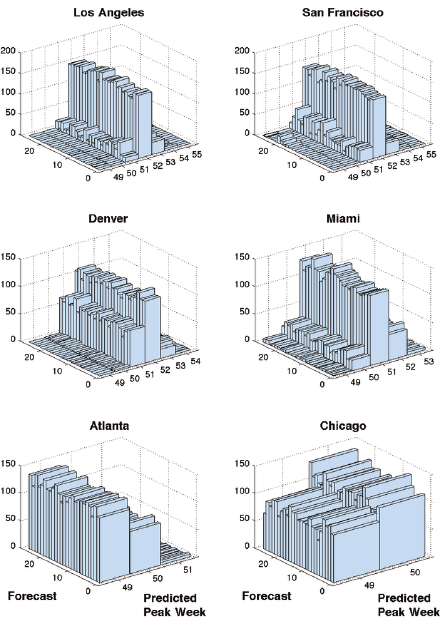}
\noindent\includegraphics[width=18pc,angle=0]{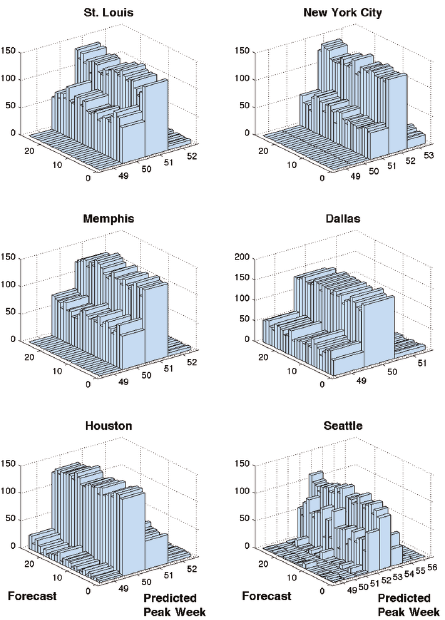}\\
\caption{Left) Histograms of the best ensemble start date trainings
  for forecasts made beginning the start of Week 51 (December 16,
  2012) for select cities.  The distributions show the ensemble spread
  among peak predictions.} 
\label{fig:select_wk50fore}
\end{figure}

The Week 50 forecast (initiated December 16, 2012) predicts peaks for
Atlanta Week 49, 1 week in the past (Figure
\ref{fig:select_wk50fore_cal}).  This is consistent with the prior
week forecast, so the forecast for Atlanta remains unchanged.  The
forecast for Chicago is for either 1 week in the past or the week just
finished (Week 49 or 50); the histograms for Chicago reveal that week
50 is favored (Figure \ref{fig:select_wk50fore}), which is a slight
shift from the prior week prediction, which was firmly on Week 49 (see
Figures \ref{fig:select_wk49fore_cal} and \ref{fig:select_wk49fore}).

Dallas has shifted slightly from a Week 49 forecast of 0-1 weeks in the future
(either Week 49 or 50) to a Week 50 forecast of 0 weeks in the future
(Week 50).  Last week Houston was predicted to peak Week 50, and this
prediction remains unchanged.  Memphis is now predicted to peak in
Week 51 (last week it was Week 50), though this really reflects a shift in
the distribution of predictions from majority Week 50 and some Week 51
(Figure \ref{fig:select_wk49fore}) to majority Week 51 and some Week
50 (Figure \ref{fig:select_wk50fore}).

The Atlantia, Chicago, Dallas, Houston and Memphis predictions all
come with high confidence (50-99\%, depending on calibration, Figure
\ref{fig:select_wk50fore_cal}).  The prediction for St. Louis is for
Week 51; this is a shift from Week 50 last week, though as for
Memphis, the histograms reveal a less marked shift.

Miami is still predicted to peak in Week 51.  New York City is
predicted to peak during Week 52, whereas last week it was Week 51.
As for Memphis and St. Louis, the shift is not a step but a shift
in the distribution of oucomes.  Both Los Angeles and San Francisco
are predicted to peak during Week 52, which is consistent with the
prior week prediction of Weeks 51-52.  Denver, too, is predicted to
peak Week 52, as before.  Seattle is predicted to peak 2-3 weeks in
the future (Weeks 52-53); this is consistent with the prior week
prediction.

All changes in predicted peak week timing are within the margin of
error of $\pm1$ week.  Table \ref{table:t1} presents the predictions
made over each of the last 4 weeks.  It can be seen that the real-time
predictions have been fairly stable over the last 4 weeks.  In
particular, Miami, Atlanta, Chicago, and St. Louis, haven't moved at
all (essentially).  Los Angeles, Denver, Dallas, and Houston have
drifted 1 week.  San Francisco, Memphis and Seattle have moved 1.5
weeks.  New York City has move a bit more, and unlike the other
cities, the predicted peak has moved closer and then farther into the
future.  Most other predictions, which have drifted, have drifted
farther into the future.

 \begin{table}[tbh]
 \caption{Summary of weekly model predictions at 12 select cities.  Weeks
   are labeled consecutively (Week 1 of 2013 is Week 53, etc.).
   Predictions were initiated at the end of Weeks 47, 48, 49 and 50.
   Figures present histogram distributions of the predictions.
   The range of prediction confidences, derived from municipal,
   regional and national calibrations, are given in parentheses.}\label{table:t1}
 \begin{center}
 \begin{tabular}{ccccrrcrc}
 \hline\hline
 City & Week 50 & Week 49 & Week 48 & Week 47\\
    & Prediction & Prediction & Prediction & Prediction \\
 \hline
  Los Angeles &52 (50-95\%) & 51-52 (35-90\%) & 51-52 (20-55\%) & 51 (15-30\%) \\
  San Francisco & 52 (35-85\%) & 51-52 (25-40\%) & 51 (30-85\%) &
  50-51 (25-60\%)  \\
  Denver & 52 (20-60\%) & 52 (20-55\%) & 51-52 (0-55\%) & 51 (10-30\%) \\
  Miami & 51 (40-60\%) & 51 (40-99\%) & 50-51 (40-55\%) & 50-51 (0-45\%) \\
  Atlanta &49 (80-99\%) & 49 (90-99\%) & 49 (80-95\%) & 49 (80-95\%) \\
  Chicago &49-50 (55-95\%) & 49 (55-95\%) & 49 (35-80\%) & 49 (35-80\%) \\
  St. Louis & 51 (85-99\%)  & 50-51 (80-99\%) & 50 (85-99\%) & 51 (30-90\%) \\
  New York City &52 (25-99\%) & 51 (25-99\%) & 52-53 (25-60\%) & 53-54
  (25-55\%)\\
  Memphis &51 (20-80\%) & 50 (20-80\%) & 50 (15-80\%) & 49-50 (15-55\%) \\
  Dallas &50 (65-90\%) & 49-50 (65-85\%) & 49 (50-75\%) & 49 (40-85\%) \\
  Houston &50 (75-90\%) & 50 (50-60\%) & 50 (50-60\%) & 49 (50-85\%) \\
  Seattle &52-53 (0-55\%) & 52-53 (5-55\%) & 51-52 (5-55\%) & 51 (5-35\%) \\
 \hline
 \end{tabular}
 \end{center}
\end{table}

\begin{figure}[tbh]
\noindent\includegraphics[width=25pc,angle=0]{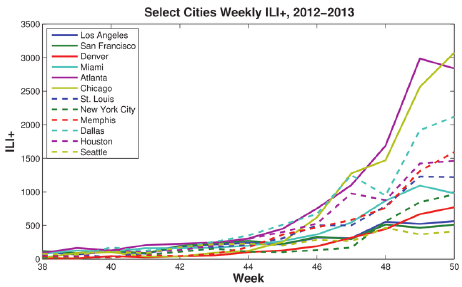}\\
\caption{Time series of Weeks 38-50 ILI+ for the 2012-2013 season.
  ILI+ is Google Flu Trends weekly municipal ILI estimates times CDC
  census division seropositive rates..} 
\label{fig:selcitytimeseries}
\end{figure}

Prediction confidences have shifted only a bit.  For instance, Seattle
has low confidence in all predictions; Atlanta and St. Louis have been
high.  Dallas, has increased through time.  Figure
\ref{fig:selcitytimeseries} shows time series of the ILI+ metric for
the 12 focus cities.  This is the evolving target, though actual
observed peak week will not be known until the outbreaks have abated.

\subsection{Week 49 Forecast}
\label{subsec:actual49}
 
\begin{figure}[tbh]
\noindent\includegraphics[width=25pc,angle=0]{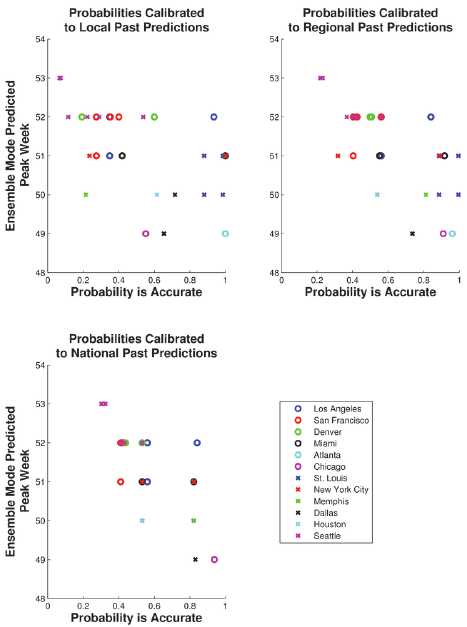}\\
\caption{Ensemble mode peak week predictions initiated December 9,
  2012, following assimilation of Week 49 observations, for 12 cities
  plotted as a function of probability/confidence calibrated from
  historical city, regional and national prediction accuracy.}
\label{fig:select_wk49fore_cal}
\end{figure}

The Week 49 forecast (initiated December 9, 2012) basically stays on
track with predictions made in prior weeks.  Atlanta and Chicago are
all predicted to be at peak ($\pm1$ week) during week 49 (Figure
\ref{fig:select_wk49fore_cal}), which is the week ending December 8,
2012.  Basically, these forecasts predict no future week higher then
that latest observed week.  Dallas is forecast to predict in 0-1
weeks.  Houston and Memphis are forecast to peak in one week (the week
ending December 15, 2012).  The calibrated confidence in these
predictions is fairly high ($>50\%$, except Memphis at the municipal scale).

\begin{figure}[tbh]
\noindent\includegraphics[width=18pc,angle=0]{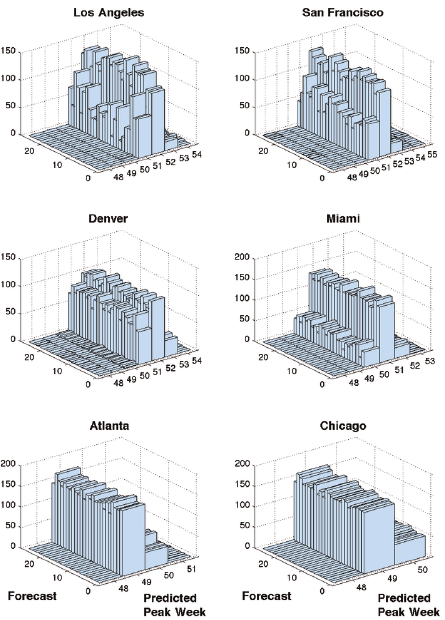}
\noindent\includegraphics[width=18pc,angle=0]{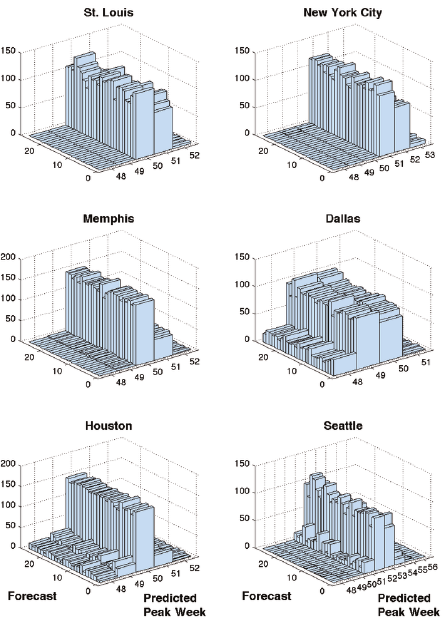}\\
\caption{Left) Histograms of the best ensemble start date trainings
  for forecasts made beginning the start of Week 50 (December 9,
  2012), following assimilation of Week 49 observations, for select
  cities.  The distributions show the ensemble spread among peak
  predictions.}
\label{fig:select_wk49fore}
\end{figure}

St. Louis is predicted to peak in 1-2 weeks.  Miami is now predicted
to peak in 2 weeks, as is New York City.  The New York City prediction
is a change of 2-3 weeks from the prior week (Week 48) prediction of
peak in 4-5 weeks.  Both Los Angeles and San Francisco are forecast to
peak in 2-3 weeks.  Denver is predicted to peak in 3 weeks (with about
$50\%$ confidence).  Seattle is predicted to peak in 3-4 weeks.
Figure \ref{fig:select_wk49fore} shows histograms of these predictions.

Most of these changes are shifts in the prediction of 1 week from the
prior week prediction, indicative of a similar tracking of outbreak
evolutions.  New York City had a larger shift of 2-3 weeks.  (Indeed,
two weeks ago, New York City had a peak predicted to be 6-7 weeks in
the future; so the forecast trajectory has shifted.)  From the
histograms, it can be seen that overall for New York City, there is
less spread among the forecasts within an ensemble--the histogram is
tightly spread between 2-3 week leads, whereas for the week prior (not
shown) there was greater spread in the predicted peak weeks.

\subsection{Week 50 Forecast -- No AH}
\label{subsec:actual50noAH}

\begin{figure}[tbh]
\noindent\includegraphics[width=25pc,angle=0]{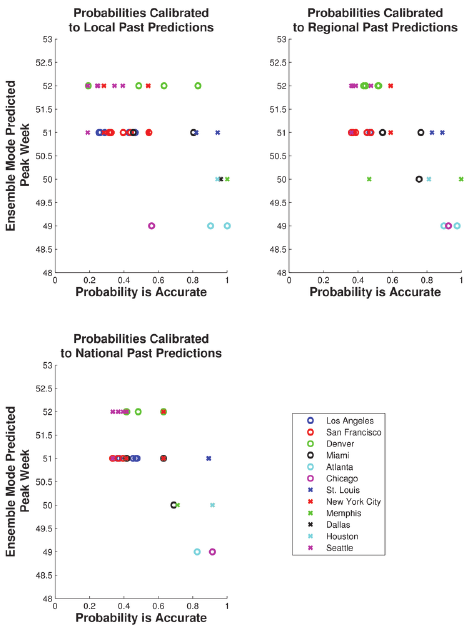}\\
\caption{Ensemble mode peak week predictions initiated December 16,
  2012, following assimilation of Week 50 observations using an SIRS
  model without AH forcing, for 12 cities plotted as a function of
  probability/confidence calibrated from historical city, regional and
  national prediction accuracy.}
\label{fig:select_wk50fore_noAH_cal}
\end{figure}

Forecasts initiated after assimilation of Week 50 observations at the
beginning of Week 51 (December 16, 2012) using an SIRS model without
absolute humidity forcing produce generally similar predictions
(Figure \ref{fig:select_wk50fore_noAH_cal}).  These similarities
include predicted outbreak peaks of Week 49 for Atlanta and Chicago,
Week 50 for Memphis and Houston, Week 51 for St. Louis, and Week 52
for Denver.  All other prediction differ by at most 1 week from those
of the model with AH forcing.

The predictions for the 12 focus cities using the SIRS model without
AH forcing are summarized in Table \ref{table:t2}).

 \begin{table}[tbh]
 \caption{Summary of weekly model predictions at 12 select cities
   using an SIRS model without absolute humidity forcing.  Weeks
   are labeled consecutively (Week 1 of 2013 is Week 53, etc.).
   Predictions were initiated at the end of Weeks 48, 49 and 50.
   Figures present histogram distributions of the predictions.
   The range of prediction confidences, derived from municipal,
   regional and national calibrations, are given in parentheses.}\label{table:t2}
 \begin{center}
 \begin{tabular}{ccccrrcrc}
 \hline\hline
 City & Week 50 & Week 49 & Week 48 \\
    & Prediction & Prediction & Prediction \\
 \hline
  Los Angeles &51 (25-50\%) & 50-51 (25-50\%) & 50 (25-50\%) \\
  San Francisco & 51 (30-60\%) & 50-51 (30-60\%) & 50 (30-50\%) \\
  Denver & 52 (50-80\%) & 51-52 (40-85\%) & 51 (40-60\%)  \\
  Miami & 50-51 (40-80\%) & 50 (10-99\%) & 50 (5-65\%) \\
  Atlanta &49 (80-99\%) & 49 (90-99\%) & 49 (25-95\%) \\
  Chicago &49 (55-95\%) & 49 (55-95\%) & 49 (25-65\%) \\
  St. Louis & 51 (80-95\%)  & 50 (80-95\%) & 50 (35-95\%) \\
  New York City &51-52 (30-60\%) & 51 (30-60\%) & 52-53 (25-60\%) \\
  Memphis &50 (45-99\%) & 50 (10-90\%) & 49-50 (15-55\%) \\
  Dallas &50 (65-90\%) & 49 (15-85\%) & 49 (40-80\%) \\
  Houston &50 (80-95\%) & 50 (60-70\%) & 49-50 (30-70\%) \\
  Seattle &51-52 (20-50\%) & 51-52 (20-45\%) & 51 (0-50\%)\\
 \hline
 \end{tabular}
 \end{center}
\end{table}

\bigskip
\textbf{Acknowledgments}

\bigskip

\myacknow

{}
{\clearpage}
\bibliographystyle{apalike} 
\bibliography{week48bib}

\end{document}